\documentstyle[aps,prl,psfig,multicol]{revtex}
\begin{document}
\newcommand{\beq}{\begin{equation}}
\newcommand{\eeq}{\end{equation}}
\draft

\title  {Tunneling spectra for quasi-one-dimensional organic
	superconductors}

\author{ N. Stefanakis}
\address{ Department of Physics, University of Crete,
	P.O. Box 2208, GR-71003, Heraklion, Crete, Greece
	}

\date{\today}
 
\maketitle

\begin{abstract}
The tunneling conductance spectra of a 
normal-metal/insulator/quasi-one-dimensional superconductor is calculated 
by using the 
Blonder-Tinkham-Klapwijk formulation. The pairing symmetry of the 
superconductor is assumed to be $p$, $d$, and $f$-wave. 
It is found that 
there is a well defined zero energy peak in electron tunneling along the 
direction parallel to the chains or normal to these 
when the transmitted quasiparticles
feel different sign of the pair potential. The actual line shape 
of the spectra is sensitive to the nodes of the pair potential on the 
Fermi surface.
\end{abstract}


\begin{multicols}{2}

\section{Introduction}
Almost $20$ years after the discovery of the organic superconductors 
\cite{jerome}, the problem of determination of their pairing state has 
not yet found a definite solution.
The critical magnetic field $H_{c2}$ exceeds the Pauli paramagnetic 
limit and indicates that the pairing symmetry is triplet \cite{lee1}.
The Knight
shift does not change between the normal and superconducting states
and it is a signature for triplet pairing state \cite{lee2}.
However the absence of Hebel-Slichter peak and the power-low decay of 
$1/T_1$ below $T_c$
\cite{takigawa}
is an indication of the presence of nodes of the pair potential on the 
Fermi surface.

The scattering theory can be used to distinguish the 
symmetry of the pair potential \cite{blonder,andreev}. 
In $d$-wave superconductors the pair potential changes sign under a
$90^o$-rotation. So under appropriate orientation of the $a$-axis within the 
$ab$ plane 
of $d$-wave superconductor the transmitted quasiparticles feel 
different sign of the pair potential. This results in the formation 
of bound states within the energy gap, which are detected as zero 
energy peaks (ZEP) in the 
conductance spectra \cite{hu,bruder,tanaka1,stefan}.

The scattering theory has also been used for the determination of the 
pairing symmetry in (TMTSF)$_2$X by Sengupta {\it et al.} \cite{sengupta}. 
However their 
calculation is restricted to the presence or absence of ZEP at the surface, 
which can not distinguish the $p$ from the $f$-wave case. In a more 
realistic calculation Tanuma {\it et al.} \cite{tanuma} 
used the extended Hubbard model 
on a quasi-one-dimensional lattice at quarter-filling, to study the 
quasiparticle states near the surface of a quasi-one dimensional 
organic superconductor, where the pairing symmetry can actually 
be distinguished from the overall line shape of the surface 
density of states (SDOS) and 
the presence or absence of ZEP.

In this paper we extend the BTK formula to calculate the tunneling 
conductance in a normal-metal/insulator/quasi-one-dimensional 
organic superconductor, where the structure of the 
Fermi surface is taken into account. 
The pairing symmetry of the
superconductor is assumed to be triplet $p$, singled  $d$, $d_{xy}$, 
and triplet $f$-wave.
In particular, we find that the ZEP appears in the 
electron tunneling along the $a$ or $b$ axis, when 
the transmitted quasiparticles experience different sign of the 
pair potential. Also the line shape of the spectra is sensitive 
to the presence or absence of nodes of the pair potential on the 
Fermi surface.
The present calculation is
useful for (TMTSF)$_2$PF$_6$, although this salt needs pressure to show
superconductivity, but is not very suitable for (TMTSF)$_2$ClO$_4$, which
is the only
material exhibiting superconductivity in the Bechgaard salts at ambient
pressure, but its
unit cell is doubled due to anion ordering \cite{yoshino}.
Also tunneling spectroscopy on the quasi two-dimensional organic 
superconductor $\kappa$-(BEDT-TTF)$_2$Cu(NCS)$_2$ has been performed by using 
scanning tunneling microscopy \cite{arai}. In this compound the conducting 
Cu(NCS)$_2$ layers run parallel to the $bc$ plane, along the $a$-axis, and 
the in-plane tunneling data strongly suggest $d$-wave pairing state. 
These features can be used to distinguish the pairing symmetry in 
quasi-one-dimensional organic superconductor.

\section{The model for the NS interface}

\begin{figure}
  \psfig{figure=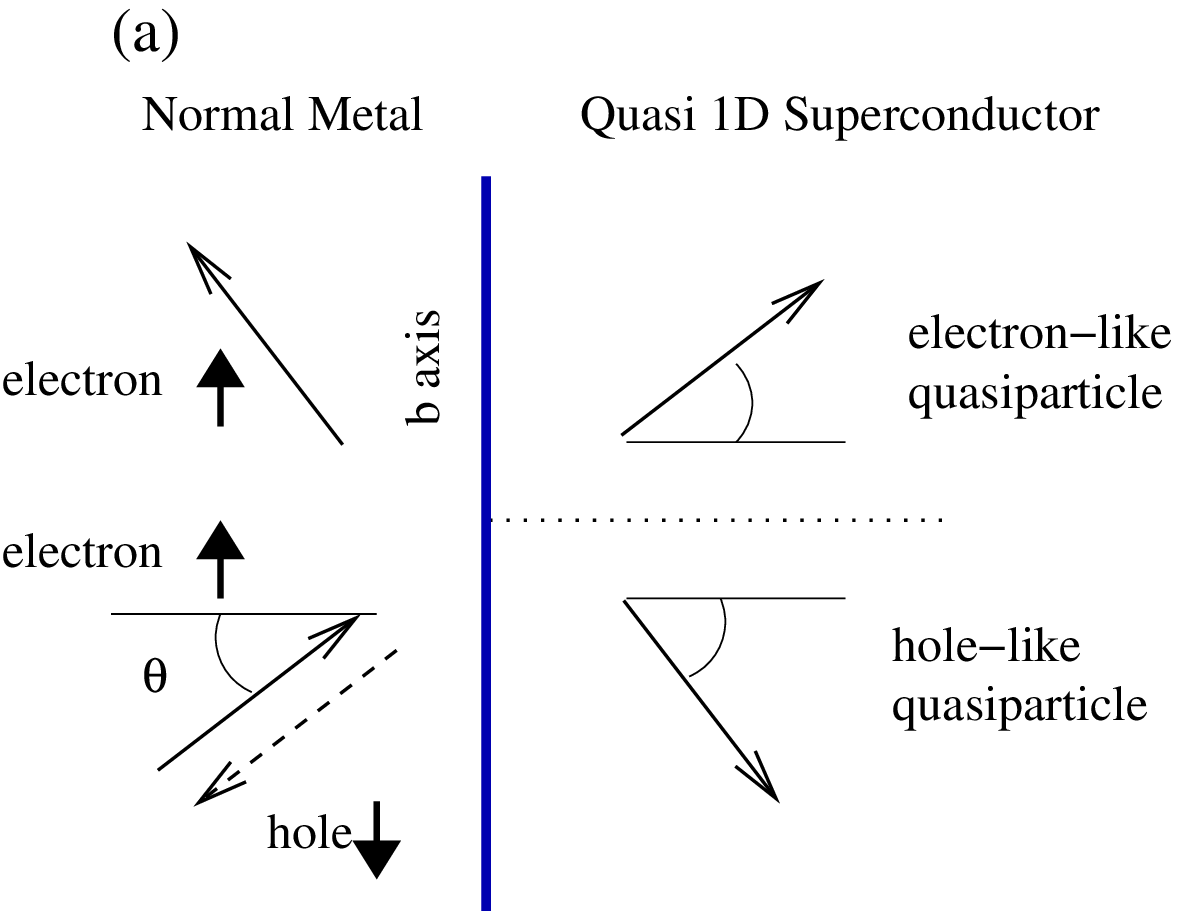,height=4.5cm,angle=0}
  \psfig{figure=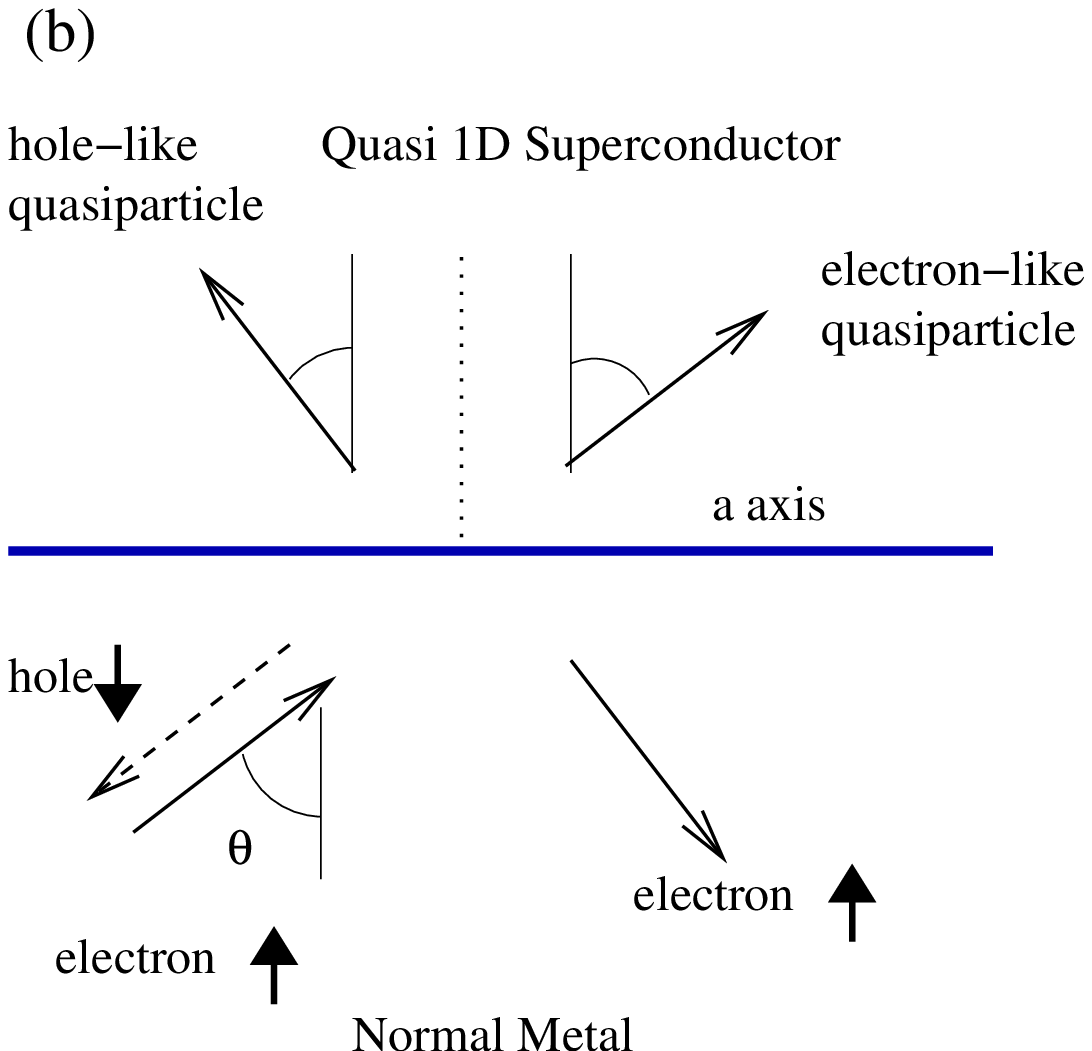,height=4.5cm,angle=0}
  \caption{
The geometry of the normal-metal/insulator/quasi-one-superconductor 
interface. 
The arrows illustrate the transmission and reflection processes at the 
interface.  
(a) The insulator (vertical line) is normal to the $a$-axis. 
(b) The insulator (horizontal line) is normal to the $b$-axis. 
}
  \label{fig1.fig}
\end{figure}

The motion of quasiparticles in inhomogeneous superconductors 
is described by the four component Bogoliubov de Gennes (BdG)
equations. 
The BdG equations are decoupled into two sets of (two component)
equations, one for the spin up electron, spin down hole 
quasiparticle wave functions ($u_{\uparrow}({\bbox r}),v_{\downarrow}({\bbox r})$) 
and the other for ($u_{\downarrow}({\bbox r}),v_{\uparrow}({\bbox r})$). 
The BdG equation for spin index 
$s(\overline{s})=\uparrow(\downarrow)$ or
$s(\overline{s})=\downarrow(\uparrow)$, read \cite{bruder}
\begin{equation}
{
\begin{array}{ll}
({\cal H}_e({\bbox r})-s\mu_B H ) u_s({\bbox r})+\int d{\bbox r'}\Delta_{s\overline{s}}({\bbox s},{\bbox x})v_{\overline{s}}({\bbox r'}) & = E_s u_s({\bbox r}) \\
\int d{\bbox r'}\Delta^{\ast}_{\overline{s}s}({\bbox s},{\bbox x})u_s({\bbox r'})-
({\cal H}_e^{\ast}({\bbox r}) -s\mu_B H ) v_{\overline{s}}({\bbox r}) & = E_sv_{\overline{s}}({\bbox r})
\end{array}.~~~\label{bdg}
}
\end{equation}
The single-particle Hamiltonian is given by ${\cal H}_e({\bbox r})=
-\hbar^2\bigtriangledown_{\bbox r}^2/2m_e+V({\bbox r})-E_F$, 
$H$ is the external magnetic field,
$E_s$ is the energy 
measured from the Fermi energy $E_F$. For a given spin projection $s$ the 
magnetic field shifts the energy by $-s\mu_B H$.
$\Delta_{s\overline{s}}({\bbox s},{\bbox x})$ is the pair potential, after a transformation 
from the position coordinates ${\bbox r},{\bbox r'}$ to the center of mass 
coordinate ${\bbox x}=({\bbox r}+{\bbox r'})/2$ 
and the relative vector ${\bbox s}={\bbox r}-{\bbox r'}$. 
After Fourier transformation the pair potential depends on the 
relative wave vector ${\bbox k}$ and ${\bbox x}$. 
In the weak coupling limit ${\bbox k}$ is 
fixed on the Fermi surface ($|{\bbox k}|=k_F$), and only its direction is 
variable. 
After applying the quasi-classical approximation, i.e., \cite{bruder}
\begin{equation}
\left(
\begin{array}{ll}
 \overline{u}_s({\bbox r}) \\
 \overline{v}_{\overline{s}}({\bbox r})
\end{array}
\right)
=e^{-i{\bbox k} \cdot {\bbox r}}
\left(
\begin{array}{ll}
  u_s({\bbox r}) \\
  v_{\overline{s}}({\bbox r})
\end{array}
\right)
,~~~\label{out}
\end{equation}
so that the
fast oscillating part, of the wave function is divided out, the BdG equations
are reduced to the Andreev equations \cite{andreev}
\begin{equation}
{
\begin{array}{ll}
E_s\overline{u}_s({\bbox r}) & = -iv_F{\bbox k} \cdot {\bbox \bigtriangledown}
\overline{u}_s({\bbox r})+
\Delta_{s\overline{s}}({\bbox k},{\bbox r})\overline{v}
_{\overline{s}}({\bbox r}) \\
E_s\overline{v}_{\overline{s}}({\bbox r}) & = iv_F{\bbox k} \cdot {\bbox \bigtriangledown}
\overline{v}_{\overline{s}}({\bbox r})+
\Delta_{\overline{s}s}^{\ast}({\bbox k},{\bbox r}) \overline{u}_{s}({\bbox r})
\end{array},~~~\label{ndrv}
}
\end{equation}
where the quantities $\overline{u}_s({\bbox r})$ and $\overline{v}_{\overline{s}}({\bbox r
})$
are electron-like and hole-like quasiparticles with spin index
$s$, and $\overline{s}$ respectively, and $v_F$ is the Fermi velocity.
We consider the normal-metal/insulator/superconductor 
junction shown in Fig. \ref{fig1.fig}.
The electron momentum parallel to the 
interface $k_{\parallel}$ is conserved.
For the interface that is normal to the $a$-axis, 
the insulator is modelled by a delta function, located at $x=0$, of the 
form $V\delta(x)$. The temperature is fixed to $0$ K.
We take the pair potential
as a step function, i.e.,
$\Delta_{s\overline{s}}(k_{\parallel},{\bbox r})=\Theta(x)\Delta_{s\overline{s}}(k_{\parallel})$.
For the geometry shown in Fig. \ref{fig1.fig}, Eqs. \ref{ndrv}
take the form
\begin{equation}
{
\begin{array}{ll}
E_s\overline{u}_s(x) & = -iv_F k_{Fx} \frac{d}{dx}
\overline{u}_s(x)+
\Delta_{s\overline{s}}(k_{\parallel})\overline{v}
_{\overline{s}}(x) \\
E_s\overline{v}_{\overline{s}}(x) & = iv_F k_{Fx} \frac{d}{dx}
\overline{v}_{\overline{s}}(x)+
\Delta_{\overline{s}s}^{\ast}(k_{\parallel}) \overline{u}_{s}(x)
\end{array}.~~~\label{ndrv1d}
}
\end{equation}

When a beam of electrons is incident from the normal metal
to the insulator, with momentum $k$, the general solution 
of Eqs. (\ref{ndrv1d}), is the two-component wave function 
$\Psi_I=(u_{\uparrow[\downarrow]},
v_{\downarrow[\uparrow]})$ 
which for $x<0$ is written as
\begin{equation}
\Psi_I=
\left(
\begin{array}{ll}
  1 \\
  0 
\end{array}
\right)
e^{i x k_{Fx} }+a_{\uparrow [\downarrow]}
\left(
\begin{array}{ll}
  0 \\
  1 
\end{array}
\right)
e^{i x k_{Fx} }+b_{\uparrow [\downarrow]}
\left(
\begin{array}{ll}
  1 \\
  0 
\end{array}
\right)
e^{-i x k_{Fx} },
~~~\label{x_}
\end{equation}
where $a_{\uparrow [\downarrow]},b_{\uparrow [\downarrow]}$, 
are the amplitudes for Andreev and normal reflection
for spin-up(-down) quasiparticles, 
$k_{Fx}$ is the $x$-component of the Fermi wave vector 
$k_F=\sqrt{\frac{2 m E_F}{\hbar^2}}$.
The wave vector of quasiparticles in the normal-metal, and
the wave vector of the electron-like and hole-like quasiparticles in 
the superconductor 
are set equal.
Since the translational symmetry holds in the $y$-axis 
direction, the momentum parallel to the interface is conserved.
Using the matching conditions of the wave function at $x=0$,
$\Psi_I(0)=\Psi_{II}(0)$ and 
$\Psi_{II}'(0)-\Psi_{I}'(0)=(2mV/\hbar^2)\Psi_I(0)$, 
the Andreev and normal reflection amplitudes
$a_{\uparrow [\downarrow]},b_{\uparrow [\downarrow]}$
for the spin-up(-down) quasiparticles are obtained as

\begin{equation}
a_{\uparrow [\downarrow]}=\frac{4n_{+}}
     {4+z_{\uparrow [\downarrow]}^2
     -z_{\uparrow [\downarrow]}^2n_{+}n_{-}
      \phi_{-}\phi_{+}^{\ast}}
,~~~\label{ra}
\end{equation}

\begin{equation}
b_{\uparrow [\downarrow]}=\frac
     {-(2iz_{\uparrow [\downarrow]}+z_{\uparrow [\downarrow]}^2)
	+(2iz_{\uparrow [\downarrow]}+z_{\uparrow [\downarrow]}^2)
     n_{+}n_{-}
      \phi_{-}\phi_{+}^{\ast}}
{4+z_{\uparrow [\downarrow]}^2
     -z_{\uparrow [\downarrow]}^2n_{+}n_{-}
      \phi_{-}\phi_{+}^{\ast}}
,~~~\label{rb}
\end{equation}
where 
$z_0=\frac{m V}{\hbar^2 k_s}$, 
$z_{\uparrow [\downarrow]}=\frac{2 z_0}{k_{Fx} }$.
The BCS coherence factors are given by 
\begin{equation}
u_{\pm}^2=[1+
      \sqrt{E^2-|\Delta_{\pm}|^2}/E]/2,
\end{equation}
\begin{equation}
v_{\pm}^2=[1-
      \sqrt{E^2-|\Delta_{\pm}|^2}/E]/2,
\end{equation}
and $n_{\pm}=v_{\pm}/u_{\pm}$.
The internal phase coming from the energy gap is given by
$\phi_{\pm} =[
\Delta_{\pm}/|\Delta_{\pm}|]$,
where $\Delta_{+}$
($\Delta_{-}$), is the 
pair potential experienced by the transmitted electron-like 
(hole-like) quasiparticle.

According to the BTK formula the conductance 
of the junction, 
$\overline{\sigma}_{\uparrow [\downarrow]}(E_s,k_{\parallel})$, 
for up(down) spin quasiparticles, 
is expressed in terms of the 
probability amplitudes
$a_{\uparrow [\downarrow]},b_{\uparrow [\downarrow]}$ as
\cite{blonder}
\begin{equation}
\overline{\sigma}_{\uparrow [\downarrow]}(E_s,k_{\parallel}) 
=1+|a_{\uparrow [\downarrow]}|^2
-|b_{\uparrow [\downarrow]}|^2
.~~~\label{ovs}
\end{equation}
The tunneling conductance, normalized by that in the normal 
state is given by 

\begin{equation}
\sigma(E)=
\sigma_{\uparrow }(E_{\uparrow })+
\sigma_{\downarrow }(E_{\downarrow })
,~~~\label{sqcharge}
\end{equation}
	
\begin{equation}
\sigma_{\uparrow [\downarrow]}(E_s)=
	\frac{1}{R_N}
\int_{-k_{\parallel}^{max}}^{k_{\parallel}^{max}}dk_{\parallel}
\overline{\sigma}_{\uparrow [\downarrow]}(E_s,k_{\parallel})
,~~~\label{sq}
\end{equation}
where
\begin{equation}
R_N=
\int_{-k_{\parallel}^{max}}^{k_{\parallel}^{max}}dk_{\parallel} [ \sigma_{N_{\uparrow}}(k_{\parallel})
	+
\sigma_{N_{\downarrow}}(k_{\parallel})]
,~~~\label{RN}
\end{equation}

\begin{equation}
\sigma_{N_{\uparrow [\downarrow]}}(k_{\parallel})
      =\frac{4\lambda_1}{4+z_{\uparrow [\downarrow]}^2}
.~~~\label{sN}
\end{equation}

\section{pairing states and Fermi surface line shape}

For the spin triplet pairing state the Cooper pairs have
spin $1$ degree of freedom.
The gap function is a $2\times 2$ symmetric matrix which in the
spin space can
be written as
\begin{equation}
\hat{\Delta}({\bbox k})= i\sigma_y
({\bbox d}\cdot \hat{\bbox \sigma} ),
\end{equation}
where $\hat{\bbox \sigma}$ denotes the Pauli matrices and
${\bbox d}$ is a 
vector which defines the axis along which
the Cooper pairs have zero spin projection. In the following
we will take
${\bbox d} \parallel \hat{\bbox a}$, i.e., parallel to the chains. 
In that case
$\Delta_{\uparrow \uparrow} =
\Delta_{\downarrow \downarrow} =0$, while
$\Delta_{\uparrow \downarrow} =
\Delta_{\downarrow \uparrow} =\Delta(k_{\parallel})$.

\begin{figure}
  \psfig{figure=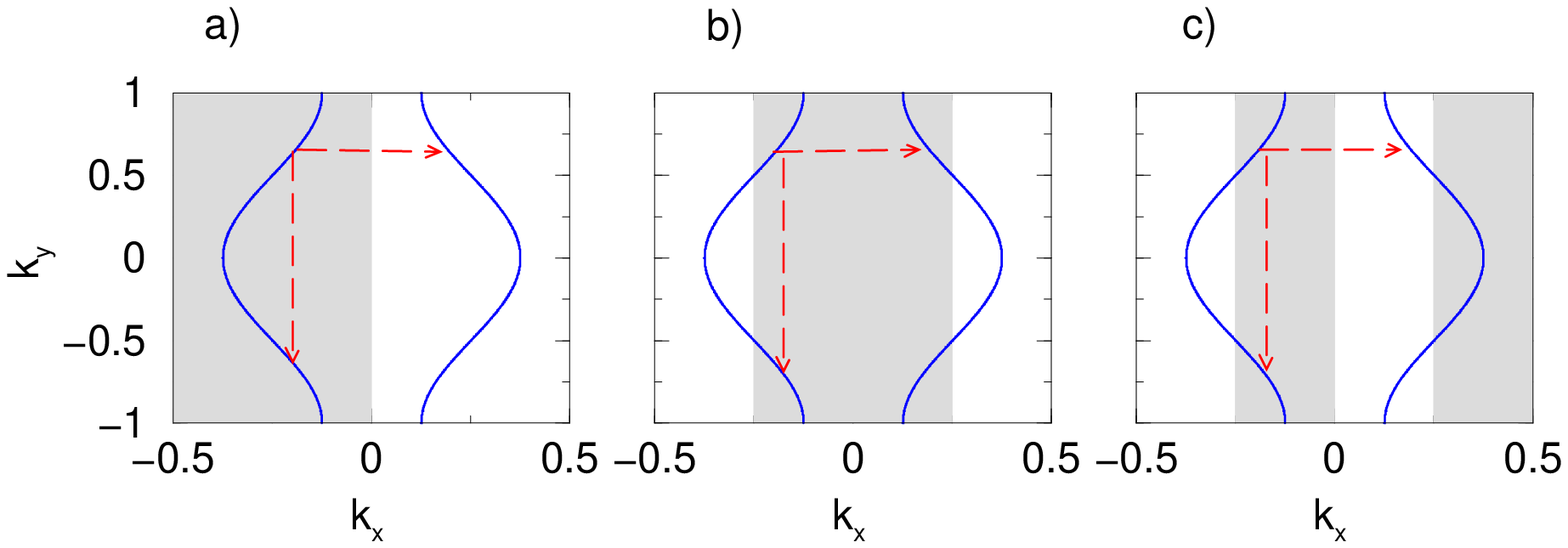,height=3.5cm,angle=0}
  \psfig{figure=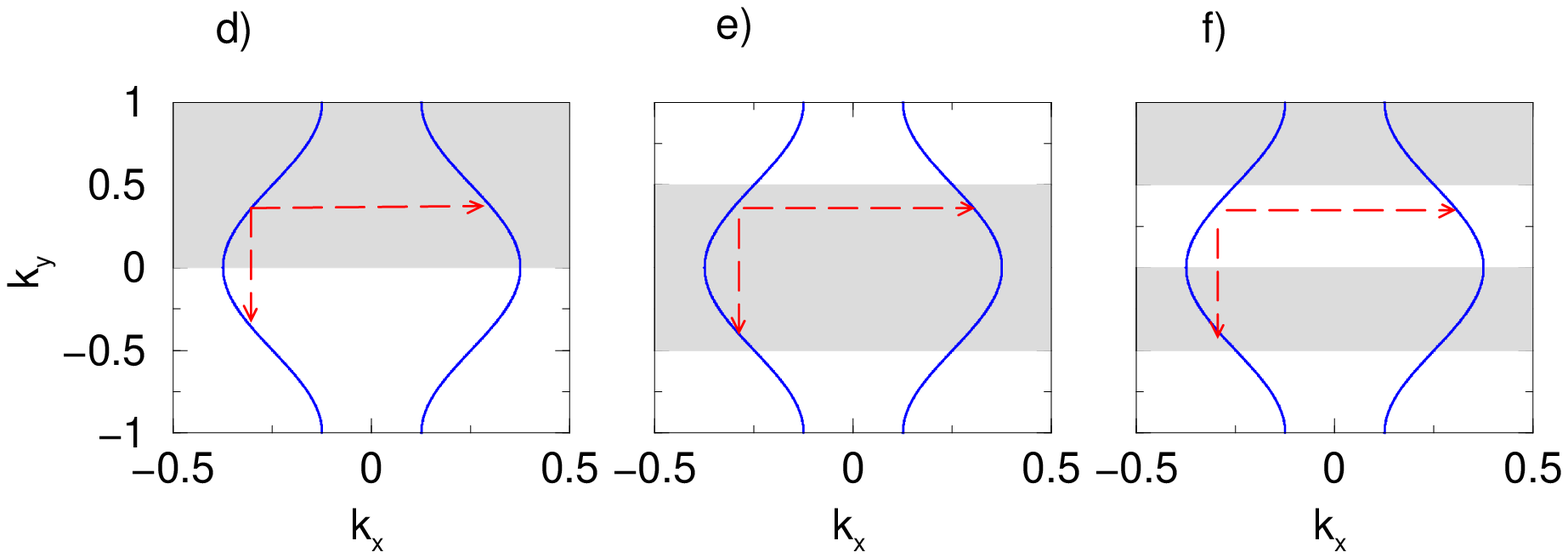,height=3.5cm,angle=0}
  \caption{
The shape of the Fermi surface (solid line) 
for the pair potentials (a) $p_x$-wave, 
(b) $d_x$-wave, (c) $f_x$-wave, (d) $p_y$-wave, (e) $d_y$-wave, (f) $f_y$-wave.
Inside the shaded(white) region the pair potential 
is negative(positive). $k_x$, $k_y$ are in units of $\pi/a$, 
$a$ is the crystal lattice spacing.
The horizontal (vertical ) arrow indicates the 
change in the momentum in the electron tunneling along the $a$ ($b$) axis. 
}
  \label{fs.fig}
\end{figure}

The Fermi surface (FS) consists of two branches and is open in the 
$k_y$ direction, as shown in Fig. \ref{fs.fig}.  

\begin{figure}
  \psfig{figure=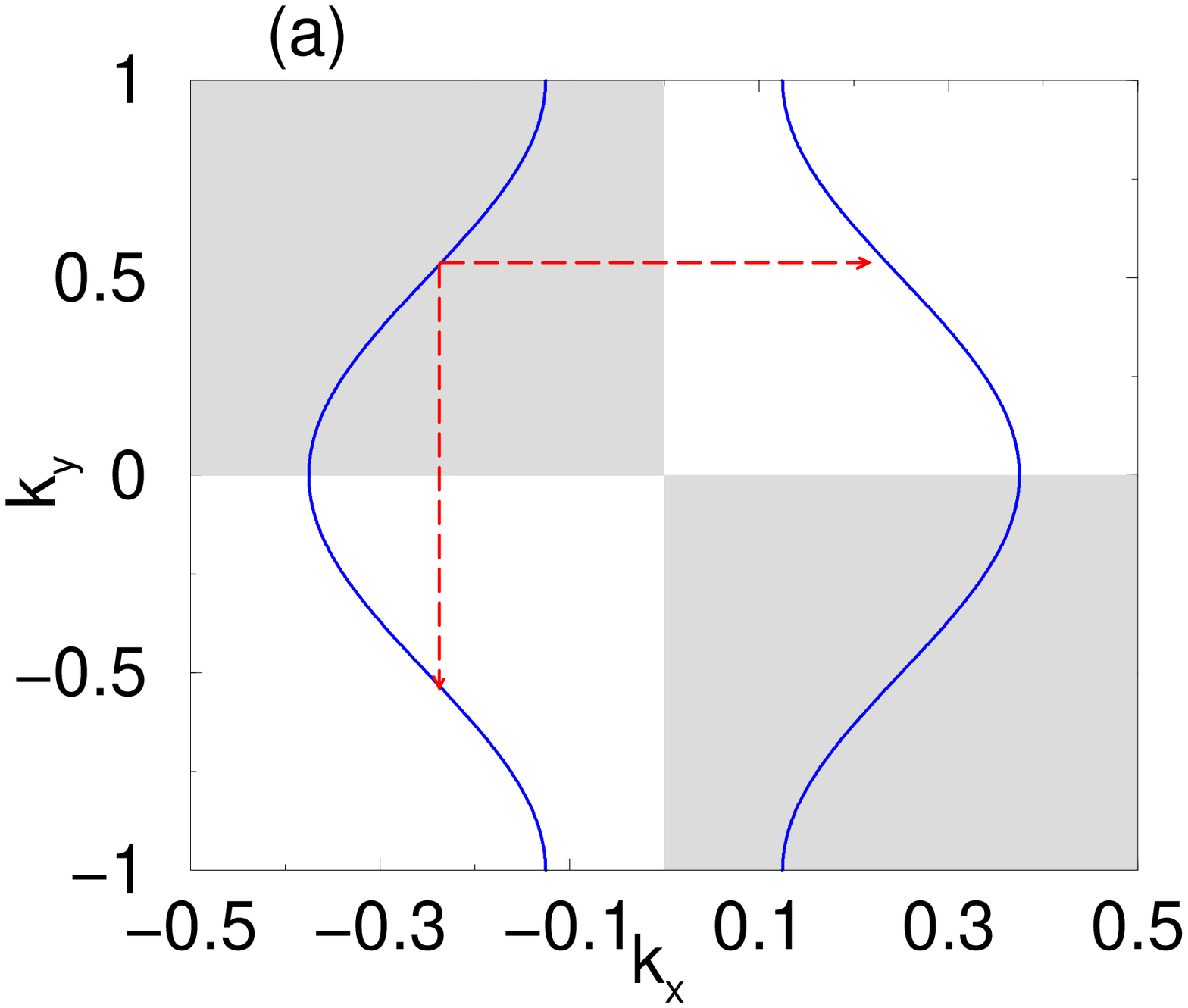,height=3.5cm,angle=0}
  \psfig{figure=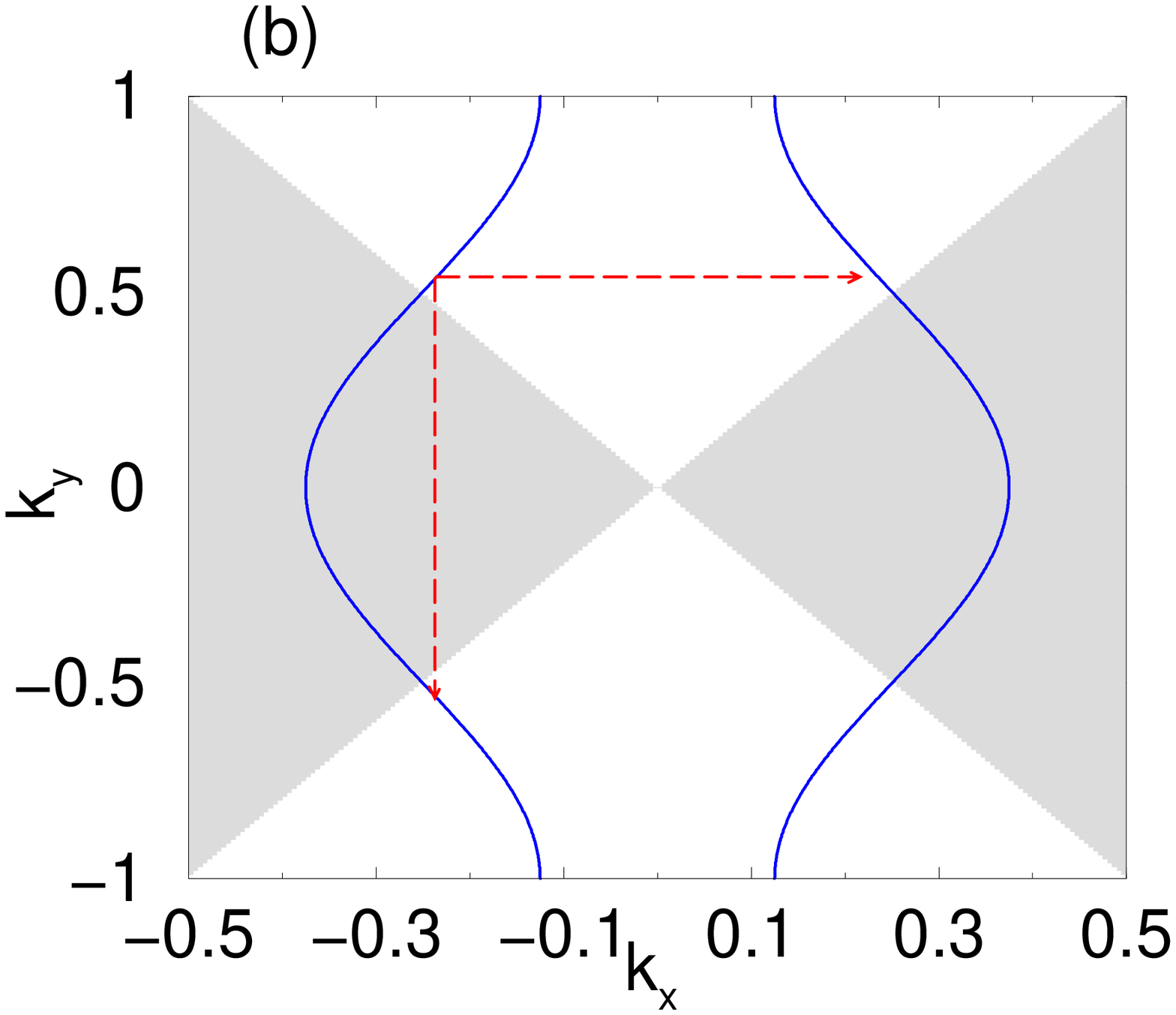,height=3.5cm,angle=0}
  \caption{
The shape of the Fermi surface (solid line) 
for the pair potentials (a) $d_{xy}$-wave , (b) $d_{x^2-y^2}$-wave.
Inside the shaded(white) region the pair potential 
is negative(positive). $k_x$,$k_y$ are in units of $\pi/a$, 
where $a$ is the crystal lattice spacing.
The horizontal (vertical ) arrow indicates the 
change in the momentum in the electron tunneling along the $a$ ($b$) axis. 
}
  \label{fsdxy.fig}
\end{figure}

We consider the following pairing cases

a) In case of $p_x$-wave, $p_y$-wave superconductor
\begin{equation}
\Delta_{p_{x}}=\Delta_0 \sin(2k_x a), \Delta_{p_{y}}=\Delta_0 \sin(k_y a).
\end{equation}
$\Delta_{p_{x}}$ changes its sign along the FS in the $k_x$ direction 
while the $\Delta_{p_{y}}$ changes its sign along the FS 
in the $k_y$ direction as seen in 
Figs. \ref{fs.fig}(a) and \ref{fs.fig}(d). 

b) In case of $d_x$-wave, $d_y$-wave superconductor
\begin{equation}
\Delta_{d_{x}}=\Delta_0 \cos(2k_x a), \Delta_{d_{y}}=\Delta_0 \cos(k_y a)
\end{equation}
$\Delta_{d_{x}}$ changes its sign along the FS in the $k_x$ direction      
while the $\Delta_{d_{y}}$ changes its sign along the FS 
in the $k_y$ direction as seen in Figs. \ref{fs.fig}(b) and \ref{fs.fig}(e).

c) In case of $f_x$-wave, $f_y$-wave superconductor
\begin{equation}
\Delta_{f_{x}}=\Delta_0 \sin(4k_x a), \Delta_{f_{y}}=\Delta_0 \sin(2k_y a)
\end{equation}
$\Delta_{f_{x}}$ changes its sign along the FS in the $k_x$ direction
while the $\Delta_{f_{y}}$ changes its sign along the FS
in the $k_y$ direction as seen in Figs. \ref{fs.fig}(c) and \ref{fs.fig}(f).

d) In case of $d_{xy}$-wave, $d_{x^2-y^2}$-wave superconductor
\begin{eqnarray}
\Delta_{d_{xy}} = & \Delta_0 \sin(2k_x a) \times \sin(k_y a), \\
\Delta_{d_{x^2-y^2}} = & \Delta_0 (\cos(2k_x a) - \cos(k_y a)),
\end{eqnarray} 
$\Delta_{d_{xy}}$, $\Delta_{d_{x^2-y^2}}$ change sign along the FS   
as seen in Figs. \ref{fsdxy.fig}(a) and \ref{fsdxy.fig}(b).

\section{Tunneling conductance characteristics} 

\begin{figure}
\centerline{
  \psfig{figure=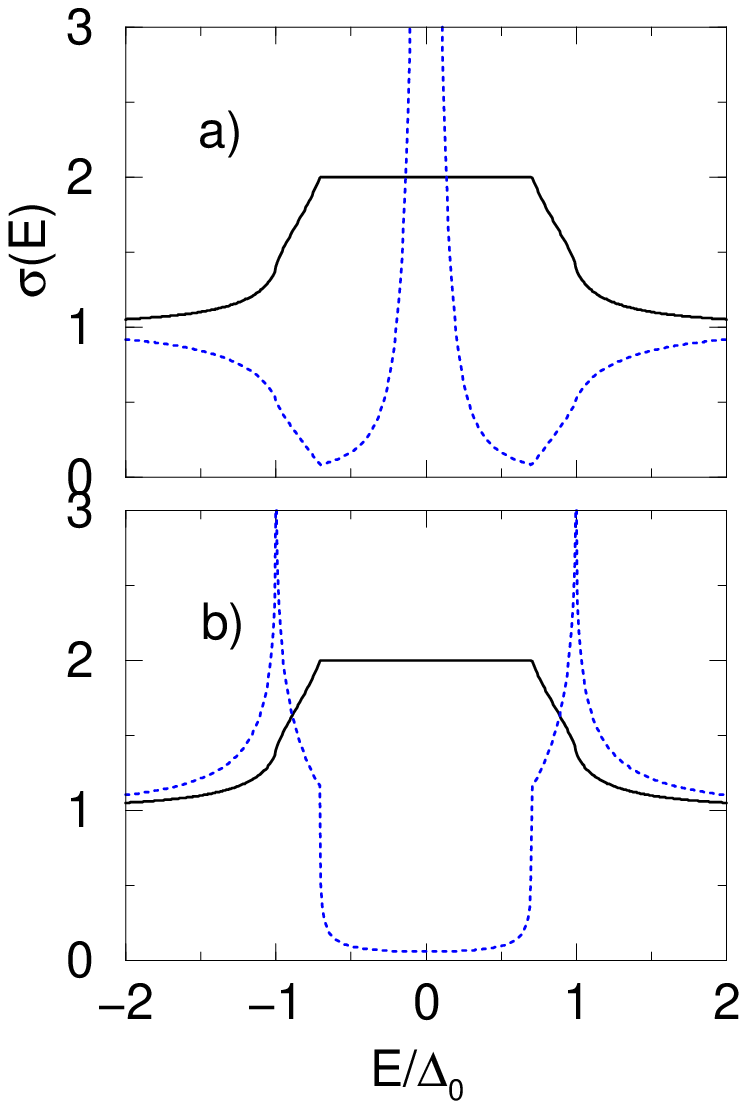,height=5cm,angle=0}
  \psfig{figure=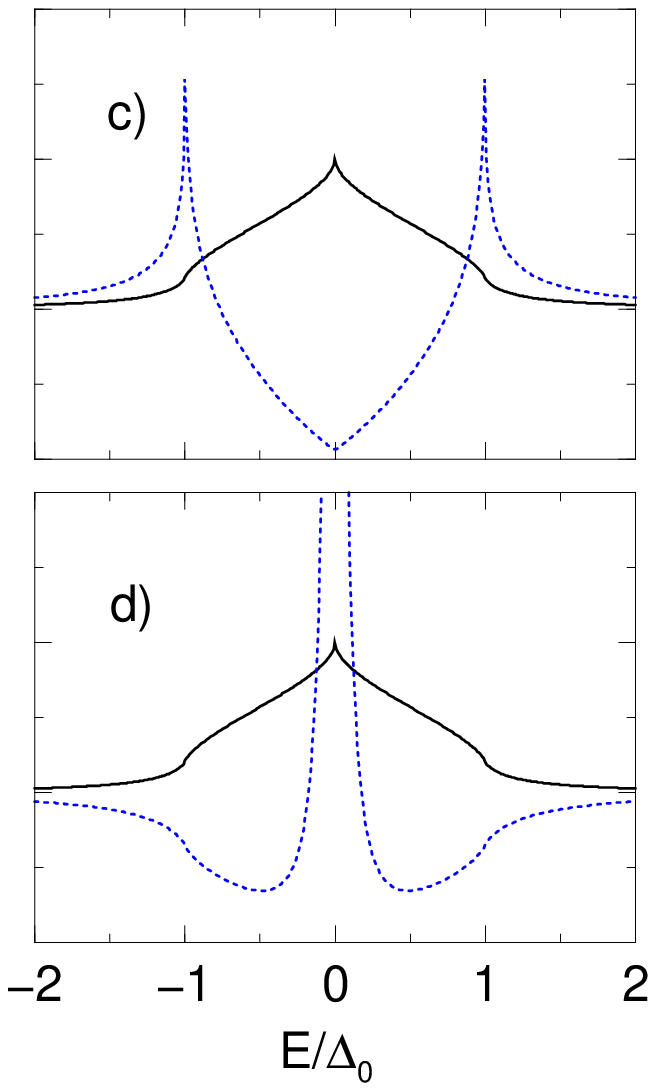,height=5cm,angle=0}
  }
  \caption{
Normalized tunneling conductance $\sigma(E)$ as a function of $E/\Delta_0$
for $z_0=0$ (solid line), $z_0=2.5$ (dotted line),
for different orientations; (a) and (c) for surface orientation normal 
to $a$-axis, 
(b) and (d) for surface orientation normal to $b$-axis.
The pairing
symmetry of the superconductor is
$p_x$ in (a) and (b) and $p_y$ in (c) and (d).
}
  \label{p.fig}
\end{figure}

\begin{figure}
\centerline{
  \psfig{figure=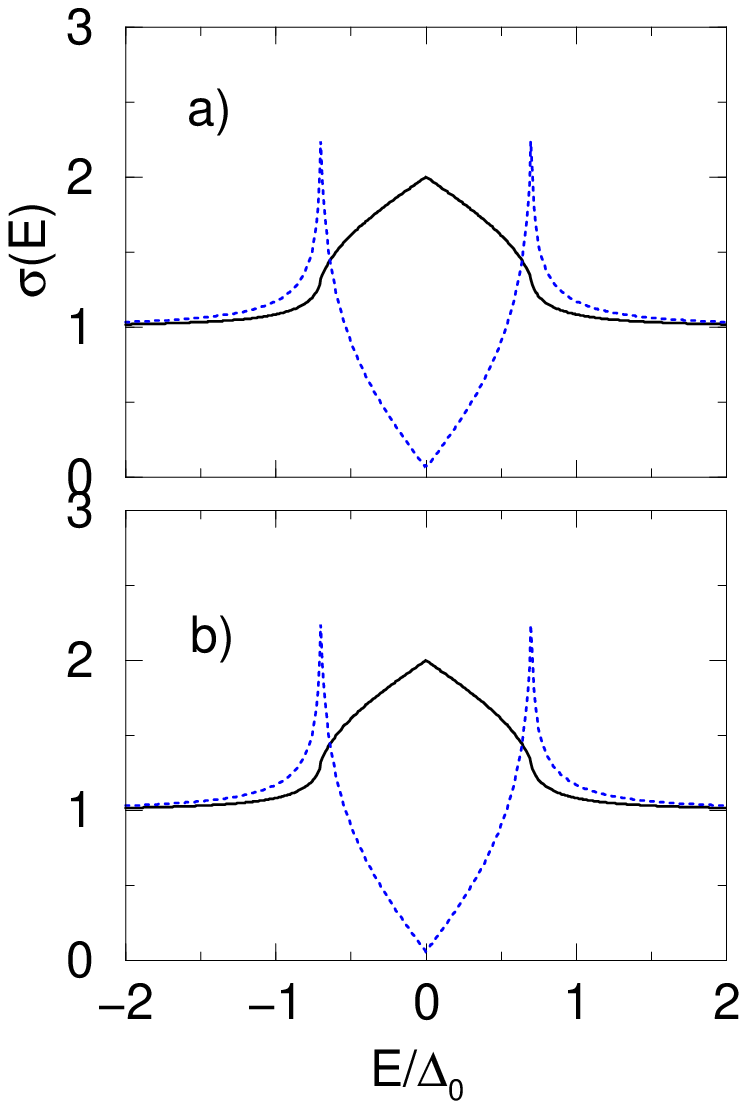,height=5cm,angle=0}
  \psfig{figure=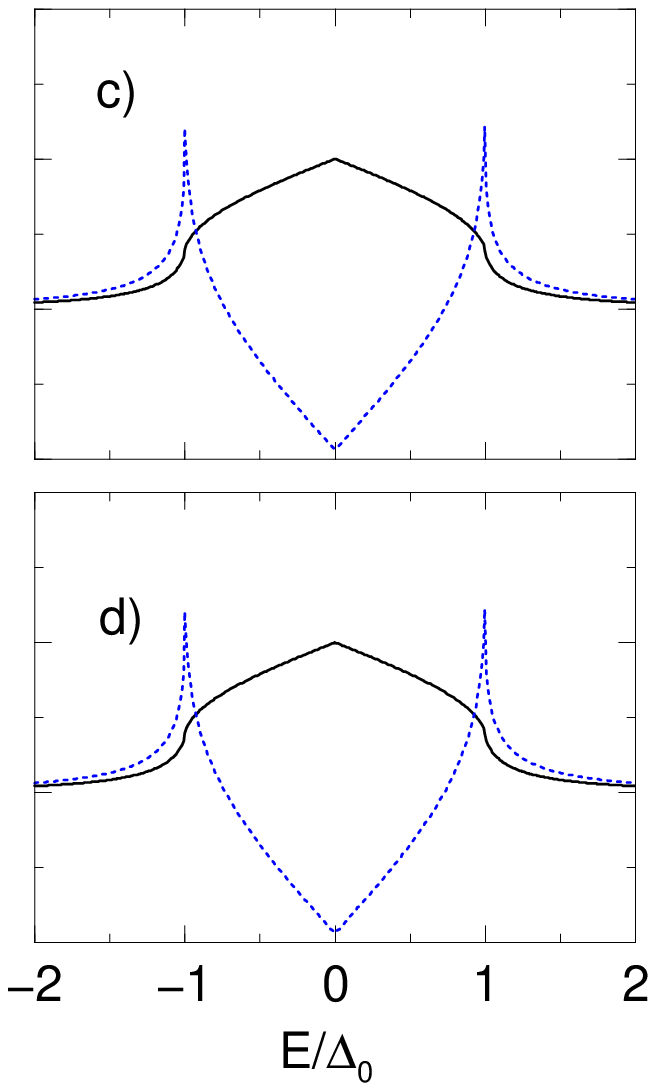,height=5cm,angle=0}
}
  \caption{
The same as in Fig. 4. The pairing
symmetry of the superconductor is
$d$-wave.
}
  \label{d.fig}
\end{figure}

\begin{figure}
\centerline{
  \psfig{figure=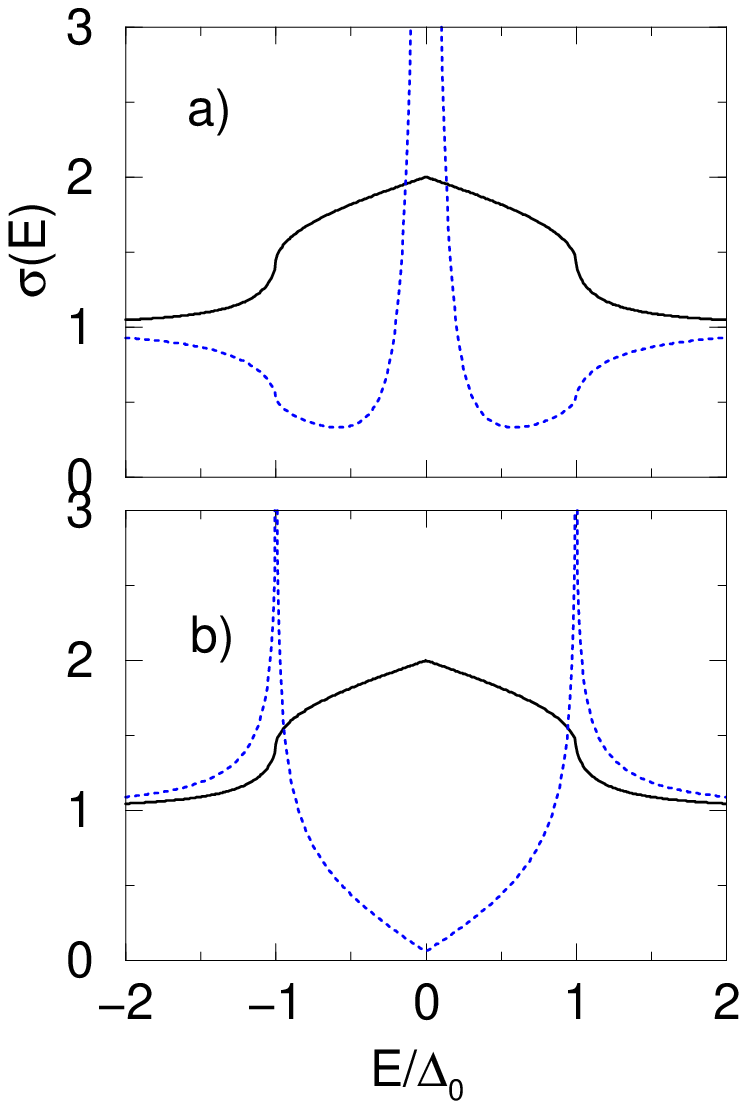,height=5cm,angle=0}
  \psfig{figure=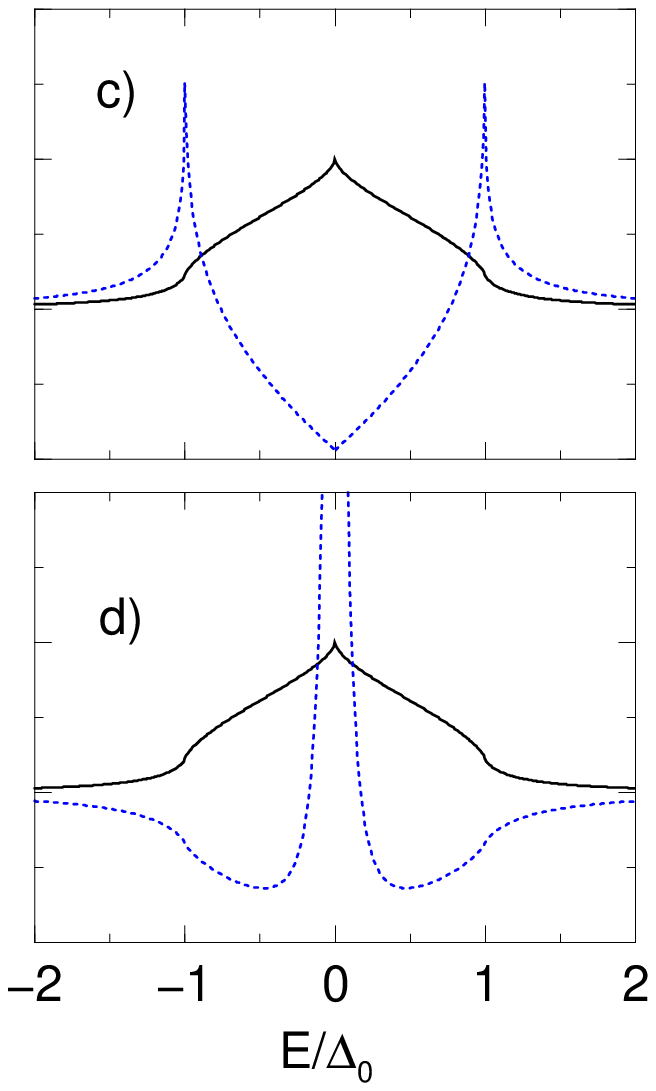,height=5cm,angle=0}}
  \caption{
The same as in Fig. 4. The pairing
symmetry of the superconductor is
$f$-wave.
}
  \label{f.fig}
\end{figure}

\begin{figure}
\centerline{
  \psfig{figure=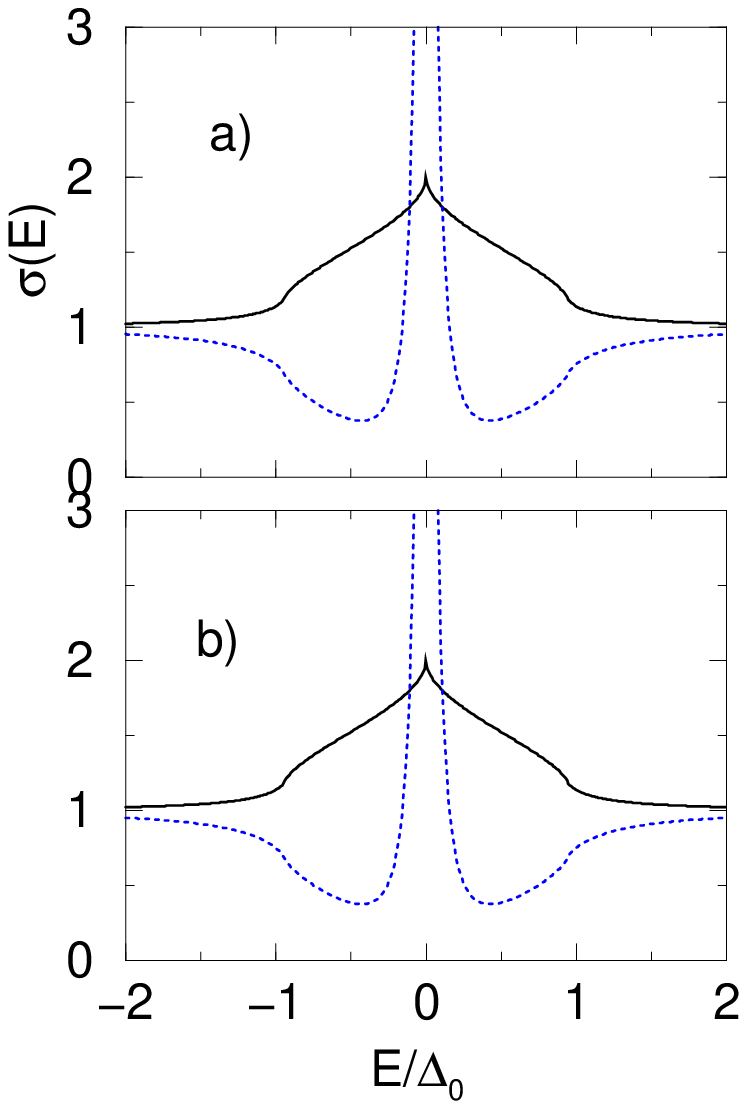,height=5cm,angle=0}
  \psfig{figure=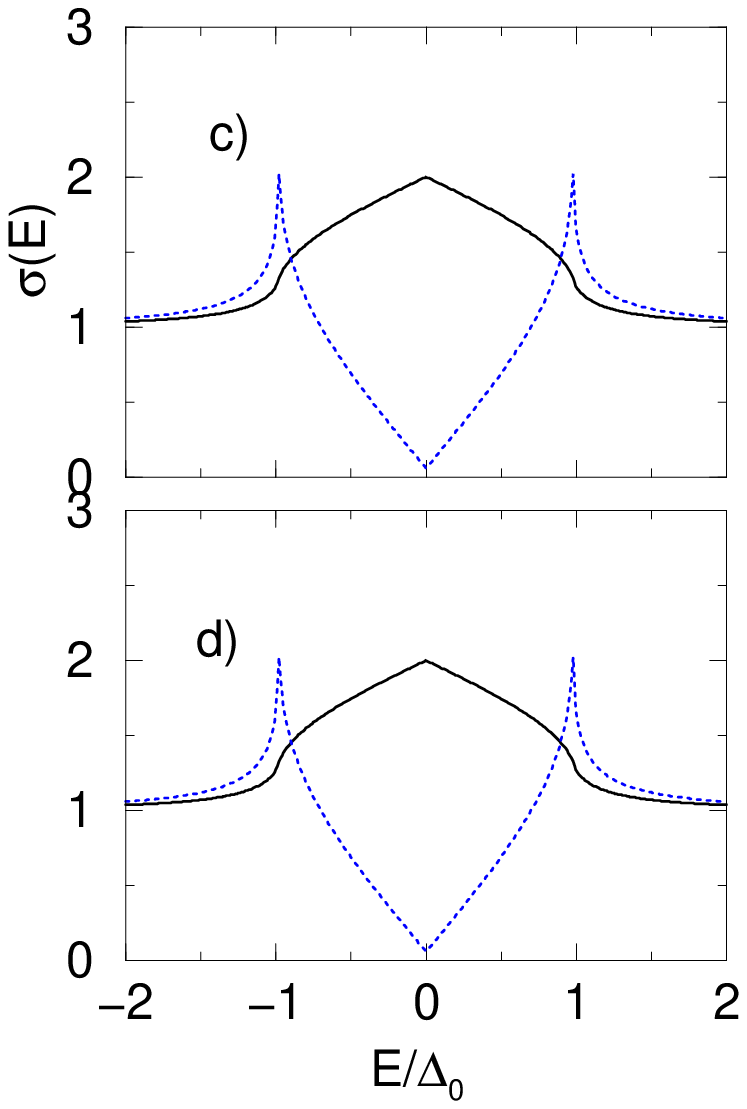,height=5cm,angle=0}
  }
  \caption{
Normalized tunneling conductance $\sigma(E)$ as a function of $E/\Delta_0$
for $z_0=0$ (solid line), $z_0=2.5$ (dotted line),
for different orientations; (a) and (c) for surface orientation normal 
to $a$-axis, 
(b) and (d) for surface orientation normal to $b$-axis.
The pairing
symmetry of the superconductor is
$d_{xy}$-wave in (a) and (b), $d_{x^2-y^2}$-wave in (c) and (d).
}
  \label{dxy.fig}
\end{figure}

In Figs. \ref{p.fig}-\ref{dxy.fig} we plot the tunneling conductance $\sigma(E)$
as a function of $E/ \Delta_0$
for various values of $z_0$, for different orientations normal to the 
$a$ or $b$ axis.
The pairing
symmetry of the superconductor is
$p_x$-wave, $p_y$-wave, in Fig. \ref{p.fig},
$d_x$-wave, $d_y$-wave, in Fig. \ref{d.fig},
$f_x$-wave, $f_y$-wave, in Fig. \ref{f.fig},
$d_{xy}$-wave, $d_{x^2-y^2}$-wave, in Fig. \ref{dxy.fig}.

The conductance peak is formed
in the electron tunneling along the $a$ or $b$ axis, when
the transmitted quasiparticles experience different sign of the
pair potential. Also the line shape of the spectra is sensitive
to the presence or absence of nodes of the pair potential on the
Fermi surface.

For the $p_x$-wave case,
when the $a$-axis of the crystal is at right angle to 
the interface, 
the horizontal line 
in Fig. \ref{fs.fig} (a), representing the scattering process,
connects points of the 
FS with opposite sign of the pair 
potential for $-\pi/a < k_y < \pi/a$. 
As a result a peak exists in the conductance spectra, 
at $E=0$ as seen in Fig. \ref{p.fig} (a) for $z_0=2.5$.
The height of the ZEP is proportional 
to the range of $k_y$  for which sign change 
occurs and it is expected to have its maximum value
when the $a$-axis of the crystal is normal to the interface,
since for this orientation the transmitted quasiparticles feel a 
different sign of the pair potential for all 
$-\pi/a < k_y < \pi/a$. 
On the other hand, when the $a$ axis is along the interface, then 
the vertical line 
in Fig. \ref{fs.fig} (a) connects points of the FS with the 
same sign of the pair
potential for $-\pi/2a < k_x < \pi/2a$, i.e.,
there is no $k_x$ for which, 
the transmitted quasiparticles feel the 
sign change of the pair potential, and no ZEP is formed as seen in 
Fig. \ref{p.fig} (b). 
For the $p_x$-wave case the nodes of the pair potential 
do not intersect the FS and the line shape of the spectra is U-like 
as in the case of the $s$-wave superconductor.

The situation is opposite in the $p_y$-wave case, where
the scattering process for the surface orientation normal to the 
$a$ ($b$)-axis, 
described by the horizontal (vertical)
line in Fig. \ref{fs.fig} (d) connects points of the 
FS with the same (opposite) sign.
As a consequence for 
the surface orientation normal to the $a$-axis, there is no  $k_y$ for which
the transmitted quasiparticles feel the
sign change of the order parameter, and no ZEP is formed, as seen in Fig. 
\ref{p.fig} (c),  
while for the interface along the $a$-axis the transmitted quasiparticles feel
the different sign of the pair potential, for all 
$-\pi /2a < k_x < \pi /2a$, and a ZEP exists as seen in Fig. 
\ref{p.fig} (d).
For the 
$p_y$-wave case the nodes of the pair potential intersect the FS and 
the spectra has a V-shaped form as in the case of $d$-wave 
superconductors.

For the $d$-wave case, and for tunneling along the $a$ or $b$-axis, 
the scattering process,
described with the horizontal or vertical
line in Figs. \ref{fs.fig}(b) and \ref{fs.fig}(e) connects points of the
FS with the same sign in all cases.
This means that the pair potential does not change 
sign and no ZEP is formed as seen in Fig. \ref{d.fig}. 
Also due to the presence of nodes of the pair potential along 
the FS, the line shape of the spectra is V-like.
 
For the $f$-wave case, 
although the pair potential has a different structure, 
the lines in Figs. \ref{fs.fig}(c) and \ref{fs.fig}(f) connect points of the 
FS with the same sign-change of the pair
potential as in the $p$-wave case.
As a result the tunneling shows ZEP as in the 
$p$-wave case, due to the sign change of the transmitted 
quasiparticles. 
However, in all cases in the $f$-wave case the pair potential 
intersects the FS and nodes are formed. 
As a consequence 
the line shape of the spectra is V-like, 
as seen in Fig. \ref{f.fig} unlike to 
$p_x$-wave case, where the pair potential is nodeless 
and the tunneling spectra has a U-shaped form.
The conclusion is that the line shape of the spectra can 
be used to distinguish the $p_x$ from the $f$-wave pairing state.
Our results are comparable to that of Tanuma {\it et al.} \cite{tanuma}
although their calculation was based on a different model, i.e.,
the extended Hubbard model
on a quasi-one dimensional lattice at quarter-filling.
 
For the $d_{xy}$-wave ($d_{x^2-y^2}$-wave) case, 
and for tunneling along the $a$ or $b$-axis, 
the scattering process,
described with the horizontal or vertical
line in Figs. \ref{fsdxy.fig}(a) and \ref{fsdxy.fig}(b) connects points of the
FS with different (the same) sign.
This means that the pair potential felt by the transmitted quasiparticles 
changes (conserves) 
sign during the scattering process and a ZEP (no ZEP) is formed 
at $E=0$ for the $d_{xy}$-wave ($d_{x^2-y^2}$-wave) case
as seen in Fig. \ref{dxy.fig}. 
Also due to the presence of nodes of the pair potential along 
the FS, the line shape of the spectra is V-like.
In addition the tunneling spectra for the $d_{xy}$-wave, for 
tunneling direction along the $a$ or $b$ axis, 
is equivalent to the tunneling spectra for $d_{x^2-y^2}$-wave
with surface orientation tilted by $\pi /4$, i.e., along the nodal 
direction. 
 
In the metallic limit ($z_0=0$), $\sigma(E)$ has the same form
for each pairing potential independently 
from the orientation.

In table \ref{tables} we summarize the results concerning the presence or 
absence of ZEP, and also the overall line shape of the tunneling spectra 
for various pairing potentials.

\section{Bound state energies}
These features are explained if we calculate the energy of the 
midgap state, which is given for large $z_0$ by the value in which 
the denominator of Eqs. (\ref{ra}) and (\ref{rb}) vanishes. 
The equation giving the 
energy peak level is written as \cite{stefan}
\begin{equation}
     \phi_{-} \phi_{+}^{\ast}n_{+}n_{-}|_{E=E_p}=1.0
.~~~\label{midgap}
\end{equation}
In the $p_x$-wave case, for surface orientation normal to $a$-axis, 
this equation has the solution
$E=0$,
for $-\pi /a < k_y < \pi /a$, since 
$n_{+}n_{-}|_{E=0}=-1$, and also the transmitted quasiparticles
feel a different sign of the pair potential, i.e., 
$\phi_{-}\phi_{+}^{\ast}|_{E=0}=-1$.
When a midgap state exists the tunneling conductance 
$\overline{\sigma}_s(E,k_y)$ is equal to $2$ 
and the peek in $\sigma(E)$
seen in Fig. \ref{p.fig}(a), is due to the normal state
conductance $R_N$ in Eq. \ref{sq}, which depends inversely on the $z_0^2$
for large $z_0$. 
For surface orientation normal to $b$-axis the range of $k_x$ for which Eq. \ref{midgap} 
has solutions collapses to zero,
and no bound states are formed. Then 
$\sigma(E)$ goes to zero as $1/z_0^2$ and there is no conductance peak.
In the $p_y$-wave case 
for surface orientation normal to $a$-axis Eq. \ref{midgap} has no solutions because 
$\phi_{-}\phi_{+}^{\ast}|_{E=0}=1$, 
while for surface orientation normal to $b$-axis 
a bound state is formed for all $k_x$ in the interval 
$-\pi /2a < k_x < \pi /2a$.
For the $d$-wave case and for tunneling along 
the $a$ or $b$ axis of the crystal, 
the 
condition $\phi_{-}\phi_{+}^{\ast}|_{E=0}=-1$ is not satisfied 
and no bound states are formed.
For the $f$-wave case 
that condition is the same as in the $p$-wave case, and thus the 
ZEP is formed as in the $p$-wave pairing state.
For the $d_{xy}$-wave ($d_{x^2-y^2}$-wave) case 
that condition $\phi_{-}\phi_{+}^{\ast}|_{E=0}=-1$ is (is not 
satisfyied for tunneling along the $a$ and $b$ axis and hence the 
ZEP (no ZEP) is formed for both tunneling directions.

\section{magnetic field effect}
\begin{figure}
  \psfig{figure=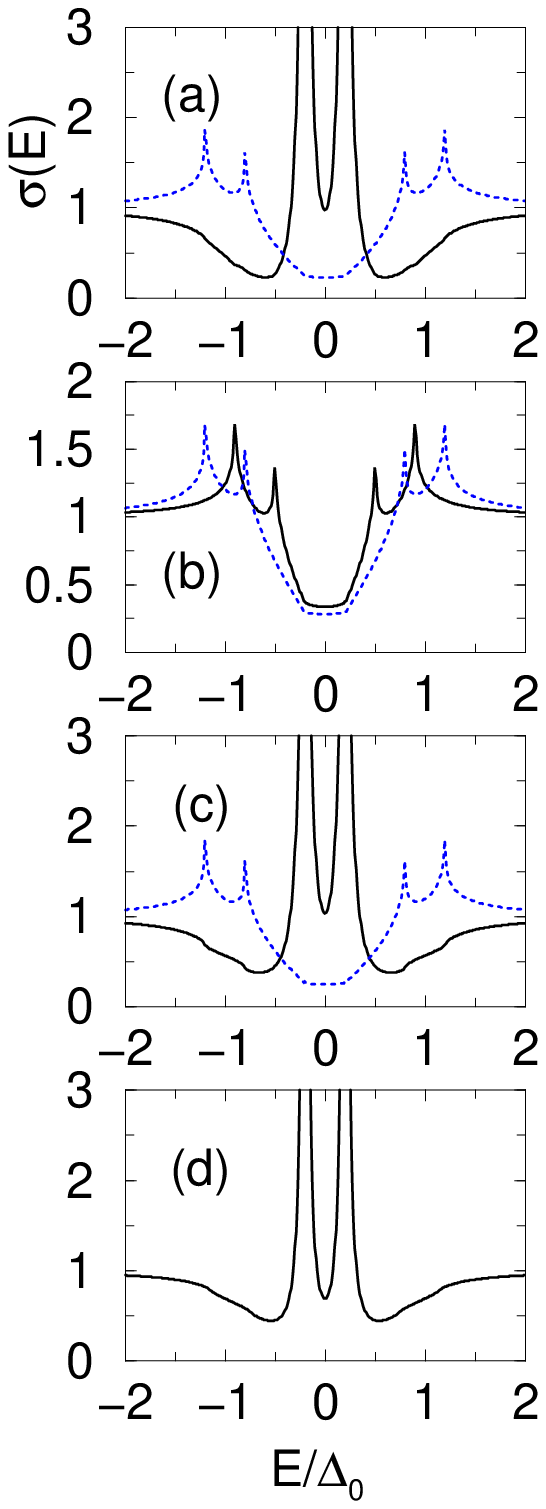,width=3.5cm,angle=0}
	
  \caption{
Tunneling conductance $\sigma (E)$ in the presence of a magnetic field
$\mu_B H/\Delta_0=0.2$ parallel to the $a$-axis of the crystal, 
as a function of the 
energy $E/\Delta_0$, for surface orientation normal to $a$-axis.
The strength of the barrier is $z=2.5$. The pairing
symmetry of the superconductor is
(a) $p_x$-wave ($p_y$-wave), solid (dotted) line, 
(b) $d_x$-wave, ($d_y$-wave), solid (dotted) line,  
(c) $f_x$-wave, ($f_y$-wave), solid (dotted) line, 
(d) $d_{xy}$-wave.
}
  \label{H.fig}
\end{figure}

In this section we describe the effect of the external magnetic
field $H$ in the direction parallel to the chains, in the spectra.
The effect of the magnetic field depends on the
spin, of the quasiparticles.
The tunneling conductance is given by
\begin{equation}
\sigma(E)= \sigma_{\uparrow}(E-\mu_B H)+
\sigma_{\downarrow}(E+\mu_B H).
\end{equation}
In Fig. \ref{H.fig}
the tunneling conductance $\sigma(E)$
is plotted for a fixed magnetic field $\mu_B H/\Delta_0=0.2$,
and
barrier strength $z_0=2.5$.
The pairing
symmetry of the superconductor is
$p$,
$d$, $f$, $d_{xy}$
respectively.

The magnetic field splits symmetrically the tunneling spectrum which is a
linear superposition of the spectra for spin up(down) quasiparticles.
The
amplitude of the splitting depends linearly on the magnetic field $H$.
For the case of $\mu_B H/\Delta_0=0.2$, seen in Fig. \ref{H.fig}
the spin up(down) part of the
spectra partially overlaps while for larger values of the magnetic
field the spin up and down branches are well
separated.
In the latter case
the right(left) branch of the spectra originates
from spin up(down) quasiparticle spectra
$\sigma_{q_{\uparrow}}(E-\mu_B H)(
\sigma_{q_{\downarrow}}(E+\mu_B H))$.

The condition for the formation of
bound states is slightly modified under the presence of a magnetic field
to $E-\mu_B H=0$, for the spin-up region, and
$E+\mu_B H=0$, for the spin-down, from the
corresponding $E=0$ in the absence of any field.
So the multiplication
of the bound states and the presence
of magnetic field results into the appearance of double peak in the
conductance spectra.

\section{Conclusions} 
We calculated the tunneling conductance in 
normal-metal/insulator/quasi-one-dimensional superconductor, 
using the BTK formalism. 
We showed that the ZEP appears in the
electron tunneling along the $a$ or $b$ axis, when
the transmitted quasiparticles experience a different sign of the
pair potential. Also the line shape of the spectra is sensitive
to the presence or absence of nodes of the pair potential on the
Fermi surface, and results to a U-shaped structure for the 
$p_x$-wave case and to a V-shaped one for the $d$ and $f$-wave cases.

The ZEP are due to the formation of bound states 
within the gap. 
The calculation of the conductance $\overline{\sigma}_s (E,k_y)$, 
for which bound state occurs, shows an enhancement at the bound state 
energy. The effect of the magnetic field in the direction parallel to the 
chains is to split linearly the ZEP. 

Throughout this paper the spatial variation of the order parameter 
near the surface, which depends on the boundary orientation, 
is ignored for simplicity.  
However, since the features presented here are intrinsic which are 
generated by the existence of surface bound states, the essential 
results do not change qualitatively. 
Also we assumed perfectly flat interfaces in the clean limit, 
so any impurity scattering and the effect of the surface roughness 
are ignored.

\begin{table}
\caption{
  Presence (Yes) or absence (No) of a zero-bias conductance peak in
  electron tunneling along the $a$ and $b$ axes
  for different symmetries of the
  pair potential. Also the overall line shape of the tunneling spectra
is shown}

\begin{tabular}{cccc}
  Symmetry &  $a$-axis ZEP & $b$-axis ZEP & line shape\\ \hline
  $p_x$ & Yes & No & U \\
  $p_y$ & No & Yes & V\\
  $d_x$ & No & No & V\\
  $d_y$ & No & No & V\\
  $f_x$ & Yes & No & V\\
  $f_y$ & No & Yes & V\\
  $d_{xy}$ & Yes & Yes & V\\
  $d_{x^2-y^2}$ & No & No & V
\end{tabular}
\label{tables}
\end{table}

\end{multicols}{2}

\begin{references}

\bibitem{jerome} D. J\'erome, A. Mazaud, M. Ribault, and K. Bechgaard,
  J. Phys. Lett. (France) {\bf 41}, L92 (1980).

\bibitem{lee1} I. J. Lee, P. M. Chaikin, and M. J. Naughton, 
Phys. Rev. B {\bf 62}, R14699 (2000).

\bibitem{lee2} I. J. Lee, S.E. Brown, W.G. Clark, M.J. Naughton, W. Kang, 
and P.M. Chaikin, Phys. Rev. Lett. {\bf 88}, 017004 (2002).

\bibitem{takigawa} M. Takigawa, H. Yasuoka, and G. Saito, J. Phys.
  Soc. Jpn. {\bf 56}, 873 (1987).

\bibitem{blonder}
G.E. Blonder, M. Tinkham, and T.M. Klapwijk, 
Phys. Rev. B {\bf 25}, 4515 (1982).

\bibitem{andreev}
A.F. Andreev, Soviet Phys. JETP {\bf 19}, 1228 (1964).

\bibitem{hu}
C.-R. Hu,
Phys. Rev. Lett. {\bf 72}, 1526 (1994).

\bibitem{bruder}
Cr. Bruder,
Phys. Rev. B {\bf 41}, 4017 (1990).

\bibitem{tanaka1}
Y. Tanaka and S. Kashiwaya, Phys. Rev. Lett. {\bf 74}, 3451 (1995).

\bibitem{stefan} N. Stefanakis,
J. Phys.: Condens. Matter {\bf 13}, 1265 (2001).

\bibitem{sengupta} K. Sengupta, I. \v{Z}uti\'c, H.-J. Kwon,
       V.M. Yakovenko, and S.D. Sarma, 
Phys. Rev. B {\bf 63}, 144351 (2001).

\bibitem{tanuma}
Y. Tanuma, K. Kuroki, Y. Tanaka, and S. Kashiwaya, 
Phys. Rev. B {\bf 64}, 214510 (2001).
 
\bibitem{yoshino} H. Yoshino, A. Oda, T. Sasaki, T, Hanajiri, J. Yamada, 
S. Nakatsuji, H. Anzai, and K. Murata, 
J. Phys. Soc. Jpn. {\bf 68}, 3142 (1999).

\bibitem{arai} T. Arai, K. Ichimura, K. Nomura, S. Takasaki, 
J. Yamada, S. Nakatsuji, and H. Anzai,  
Phys. Rev. B {\bf 63}, 104518 (2001).

\end{references}
\end{document}